\documentclass[a4paper,11pt]{article}
\usepackage{pos}
\usepackage{orcidlink}

\title{QCD in strong magnetic fields: fluctuations of conserved charges and equation of state}
\ShortTitle{QCD in strong magnetic fields: conserved charges and EoS}

\author[a]{Heng-Tong Ding~\orcidlink{0000-0003-0590-081X}}
\author[a]{Jin-Biao Gu~\orcidlink{0000-0001-9655-0186}}
\author*[a]{Arpith Kumar~\orcidlink{0000-0002-5887-3803}}
\author[a]{Sheng-Tai Li~\orcidlink{0000-0003-0059-7778}}
\affiliation[a]{Key Laboratory of Quark and Lepton Physics (MOE) and Institute of Particle Physics, \\
Central China Normal University, Wuhan 430079, China}
\emailAdd{hengtong.ding@ccnu.edu.cn} \emailAdd{jinbiaogu@mails.ccnu.edu.cn}\emailAdd{arpithk@ccnu.edu.cn}
\emailAdd{lishengtai@mails.ccnu.edu.cn}

\FullConference{The 42nd International Symposium on Lattice Field Theory (LATTICE2025)\\
2-8 November 2025\\
Tata Institute of Fundamental Research, Mumbai, India\\}

\abstract{
We present continuum-estimated (2+1)-flavor lattice QCD results for second-order fluctuations of conserved charges and the leading-order equation of state in the presence of strong magnetic fields at nonzero baryon chemical potential, using the HISQ action at the physical pion mass. The baryon--electric charge correlation $\chi^{\rm BQ}_{11}$ exhibits striking sensitivity to the magnetic field: $R_{cp}$-like double ratios $\chi^{\rm BQ}_{11}/\chi^{\rm Q}_{2}$ and $\chi^{\rm BQ}_{11}/\chi^{\rm QS}_{11}$ reach enhancements of $\sim2$ and $\sim2.25$ at $eB \simeq 8M_\pi^2$ along the transition line, establishing $\chi^{\rm BQ}_{11}$ as a magnetometer of QCD. To bridge theoretical predictions and experimental observations, we construct HRG-based proxy observables and apply systematic kinematic cuts emulating STAR and ALICE detector acceptances, which retain $\sim80\%$ of the lattice QCD magnetic sensitivity. Extending to the QCD equation of state under strangeness neutrality and isospin asymmetry, we determine the chemical potential ratio $q_1\equiv(\mu_{\rm Q}/\mu_{\rm B})_{\rm LO}$ and the pressure coefficient $P_2$ for magnetic field strengths up to $eB \simeq 0.8~{\rm GeV}^2 \sim 45 M_{\pi}^2$. The results reveal temperature-band crossings, hierarchy reversals, and non-monotonic structures driven by the nontrivial interplay between thermal and magnetic effects.

}

\begin{document}
\maketitle

\section{Introduction}
\label{sec:intro}

Strong magnetic fields are expected in diverse environments: the early Universe, magnetars, and the laboratory setting of relativistic heavy-ion collisions (HIC)~\cite{Kharzeev:2007jp}. In off-central HICs, the produced field strength $eB$ can reach the order of $\Lambda_{\rm QCD}^2$, a characteristic scale of the strong interaction. Model estimates suggest early-stage strengths $eB\sim5~M_\pi^2$ at the RHIC and $eB\sim70~M_\pi^2$ at the LHC for $^{208}_{82}$Pb$\big/ ~^{197}_{79}$Au nuclei collisions~\cite{Deng:2012pc,Skokov:2009qp}. Although transient and rapidly decaying in vacuum, such fields can be sustained long enough by the electrical conductivity and magnetic response of the collision medium~\cite{Astrakhantsev:2019zkr,Bali:2020bcn,Ding:2016hua}, governed by magnetohydrodynamics~\cite{Huang:2022qdn}. If sufficiently sustained, they can profoundly alter the QCD equation of state (EoS) and thereby modify the hydrodynamic evolution of the produced matter. These fields can also manifest as striking macroscopic phenomena, most prominently the chiral magnetic effect~\cite{Kharzeev:2007jp,Fukushima:2008xe}. The quest to detect such imprints in final-state observables has spurred intensive theoretical and experimental efforts. Over the past decade, lattice QCD studies in magnetic backgrounds have revealed significant modifications in key QCD properties, including isospin symmetry breaking, lowering of the transition temperature and the emergence of inverse magnetic catalysis~\cite{Bali:2011qj,Bali:2012zg,Endrodi:2019zrl, Ding:2025pbu}. However, most such studies focus on chiral condensates, which are not directly accessible in experiments.

Fluctuations of and correlations among net baryon number (B), electric charge (Q), and strangeness (S) are accessible both in theory and in experiment and serve as powerful tools for probing the QCD phase structure~\cite{HotQCD:2012fhj,STAR:2019ans, ALICE:2025mkk} and for constructing the QCD EoS at finite baryon density~\cite{Bazavov:2017dus}. In external magnetic backgrounds, however, studies of these fluctuations remain scarce and are largely confined to effective models~\cite{Adhikari:2024bfa}. First-principles lattice QCD calculations are indispensable for establishing model-independent benchmarks. The first lattice study in this direction employed heavier-than-physical pion mass ($M_\pi\simeq 220~\rm MeV$) at a single lattice spacing~\cite{Ding:2021cwv} and has since been extended to the physical pion mass~\cite{Ding:2023bft,Ding:2025jfz}. These conserved-charge fluctuations have furthermore been used to construct the QCD EoS in a magnetic background at nonzero density in $(2+1+1)$-flavor QCD at finite lattice spacing~\cite{Borsanyi:2023buy}, at imaginary chemical potential~\cite{Astrakhantsev:2024mat,MarquesValois:2025nzo}, and via continuum-estimated leading-order Taylor coefficients~\cite{Kumar:2025ikm, Ding:2025nyh}.

In this proceedings, we report second-order lattice QCD results for the baryon--electric charge correlation and the ratio $\mu_{\rm Q}/\mu_{\rm B}$ in nonzero magnetic fields with physical pions~\cite{Ding:2023bft,Ding:2025jfz}. To bridge these results with experiment, we employ Hadron Resonance Gas (HRG) model-based proxies and implement systematic kinematic cuts that account for detector acceptance limitations. Furthermore, under strangeness neutrality and isospin asymmetry conditions relevant for heavy-ion collisions, we present lattice QCD results for the leading-order chemical potential ratio $q_1\equiv(\mu_{\rm Q}/\mu_{\rm B})_{\rm LO}$ and the pressure coefficient $P_2$, revealing non-monotonic structures and temperature-hierarchy reversals at strong magnetic fields~\cite{Ding:2025nyh}.

\section{Thermodynamics in conserved-charge basis and HRG model }
\label{sec:thm}

\begin{figure}[htbp]
\centering
\includegraphics[width=0.45\textwidth,clip]{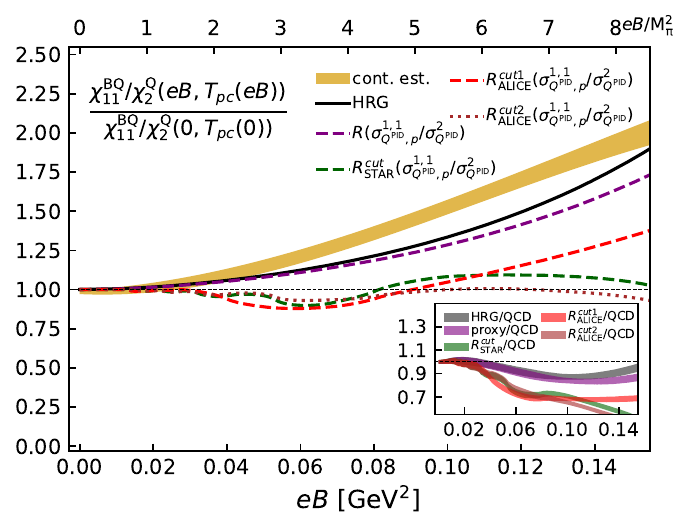}
\includegraphics[width=0.45\textwidth,clip]{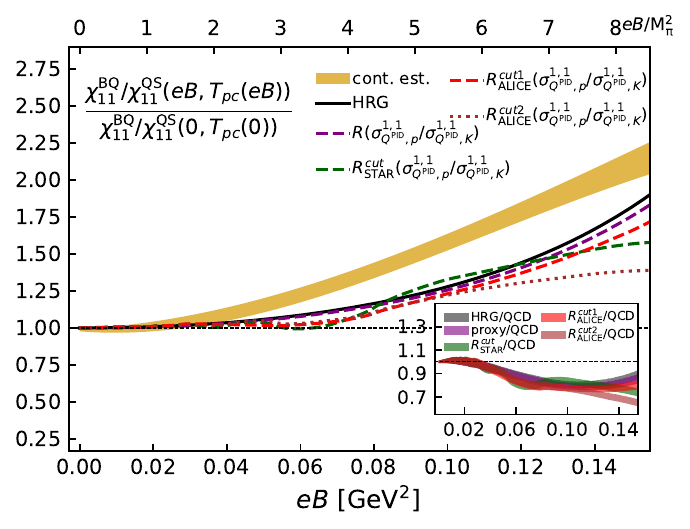}
\caption{$R_{cp}$-like ratio $R\left(\chi^{\rm BQ}_{11} /\chi^{\rm Q}_{2}\right)$ (left) and $R\left(\chi^{\rm BQ}_{11} /\chi^{\rm QS}_{11}\right)$ (right) along the transition line $T_{pc}(eB)$. Bands represent lattice data, while solid and broken lines represent HRG results.  Plots are taken from Ref.~\cite{Ding:2025jfz}.}
\label{fig:ratio_BQ-Tpc}       
\end{figure}

In a magnetized thermal medium, the QCD pressure $ P = (T/V)\ln \mathcal{Z} (eB, T, V, {\mu})$ admits a Taylor expansion in the conserved-charge chemical potentials around vanishing values:
\begin{equation}
    \hat{P} \equiv \frac{P}{T^4} =\sum_{ijk}\frac{1}{i!j!k!}~ \chi^{{\rm B} {\rm Q} {\rm S}}_{ijk}~ \hat{\mu}^{i}_{{\rm B}} \hat{\mu}^{j}_{{\rm Q}} \hat{\mu}^{k}_{{\rm S}}\,
\end{equation}
with generalized susceptibilities $\chi^{\rm BQS}_{ijk}$ defined as     
\begin{equation}
    \chi^{\rm BQS}_{ijk} =\frac{1}{VT^3} \left( \frac{\partial}{\partial  \hat{\mu}_{\rm B}} \right)^i \left( \frac{\partial}{\partial  \hat{\mu}_{\rm Q}} \right)^j \left( \frac{\partial}{\partial  \hat{\mu}_{\rm S}} \right)^k \ln \mathcal{Z}\Big|_{\hat{\mu}_{\rm B,Q,S}=0}\,,
    \label{eq:suscp_uds}
\end{equation}
where the leading-order coefficients with $i+j+k=2$ are the fluctuations of and correlations among conserved charges. Continuum estimates of these susceptibilities and pressure-related observables are obtained from lattice QCD on $32^3\times 8$ and $48^3\times 12$ lattices with the HISQ action at physical pion mass, for temperatures around $T_{pc}$ and magnetic field strengths up to $0.8~{\rm GeV}^2 \sim 45M_{\pi}^2$. The constant U(1) magnetic background is implemented as in Refs.~\cite{Ding:2020hxw,Ding:2021cwv,Ding:2025jfz}.

In the HRG framework, charged hadrons in an external magnetic field (along the $y$-direction) are subject to Landau quantization, and the transverse phase space measure is modified as $\int \mathrm{d}^3 \boldsymbol{p} = \int \left|q_R\right| B~ \mathrm{d} l ~\mathrm{d} p_y ~\mathrm{d} \phi_p$. The pressure contribution of each charged resonance is then given by~\cite{Ding:2021cwv,Ding:2023bft,Ding:2025jfz, Bhattacharyya:2015pra}:
\begin{equation}
    \frac{P^c_{R, ~\rm HRG}}{T^4}=\frac{\left|q_R\right| B}{(2 \pi)^3 T^3} \sum_{s_y
    =-s_R}^{s_R} \sum_{l=0}^{\infty} \int_{0}^{\infty}  \mathrm{d} p_y \int_{0}^{2\pi}  \mathrm{d} \phi_p   ~ \sum_{k=1}^{\infty} {(\pm 1)^{k+1}} \frac{e^{-k\left(E^c_R-\mu_R\right)/T}}{k} ,
    \label{eq:pcR}
\end{equation}
with Landau-quantized energies $E^c_R(p_y,l,s_y)=\sqrt{m_R^2+ p_y^2+2\left|q_R\right| B(l+1/2-s_y)}$ and $\mu_R=\mu_{\rm B} {\rm B}_R+ \mu_{\rm Q}{\rm Q}_R+\mu_{\rm S}{\rm S}_R$.

\section{Baryon electric charge correlations as QCD magnetometer}
\label{sec:results_chiBQ}

The baryon--electric charge correlation, $\chi^{\rm BQ}_{11}\equiv\chi^{\rm BQ}_{11}(T,eB)$, exhibits a striking sensitivity to magnetic field strength, as demonstrated by recent lattice QCD computations~\cite{Ding:2023bft,Ding:2025jfz}. Within the HRG model, the dominant contribution to the magnetic enhancement of $\chi^{\rm BQ}_{11}$ originates from the doubly charged $\Delta^{++}(1232)$. Since these $\Delta$ baryons decay rapidly to stable particles, $\Delta^{++}\to p+\pi^+$, their fluctuation contributions in experiments are accessed through final-state hadrons: protons ($p$), pions ($\pi$), and kaons ($K$). Accordingly, to align with experiments, we construct proxy observables by mapping net-conserved charges onto detectable species, $\text{net-\{B,Q,S\}} \to ~\{\tilde{p},~{Q}^{\rm PID}\equiv \tilde{\pi}^+ + \tilde{p}+\tilde{K}^+,~\tilde{K}^+\}$~\cite{Ding:2023bft,Ding:2025jfz}:
\begin{equation}
\sigma_{p,Q^{\rm PID},K}^{i,j,k}=\sum_R (\omega_{R\rightarrow \tilde{p}})^i~ (\omega_{R\rightarrow {Q}^{\rm PID}})^j ~\left(\omega_{R\rightarrow \tilde{K}}\right)^k \times I_2^{R}, 
\label{eq:proxies}
\end{equation}
where $I_2^{R} = {\partial^2 (P_R / T^4)}/{\partial \hat{\mu}_R^2}~\big|_{\hat{\mu}_{\mathrm{B}, \mathrm{Q}, \mathrm{S}}=0}$. To account for detector acceptances, we impose kinematic cuts on transverse momentum $p_T$ and pseudo-rapidity $\eta$ via a Heaviside step function $\Theta$, restricting the momentum-
space integration $\left( \int \mathrm{d} p_y \int \mathrm{d} \phi_p  \right)\times ~ \Theta(p_{T_{\min}}, p_{T_{\max}}, \eta_{\min}, \eta_{\max})$. We define cut weights $ \omega^{cuts}_{\pi,K,p}  = I_2^{R\in\{{\pi,K,p} \},~cuts}/I_2^{R\in\{{\pi,K,p} \}} $~\cite{Ding:2025jfz,Ding:2025dvn} that modify Eq. (\ref{eq:proxies}) in the following manner:
\begin{equation}
    \sigma_{p,Q^{\rm PID},K}^{i,j,k} :\omega_{R\rightarrow~\tilde{p},\tilde{\pi},\tilde{K}}~~~ \longrightarrow ~~~ \sigma_{p,Q^{\rm PID},K}^{i,j,k;~ cuts} :\omega_{R\rightarrow~\tilde{p},\tilde{\pi},\tilde{K}}~\omega^{ cuts}_{p,\pi,K}.
\end{equation}

Since magnetic-field effects in HICs are expected to vary with centrality, we define $\chi^{\rm BQ}_{11}$-based $R_{cp}$-like (central-to-peripheral) double ratios,
\begin{equation}
    R(\mathcal{O}) \equiv \mathcal{O}\left( eB,T_{pc}(eB)\right)~\big/ ~\mathcal{O}\left(eB=0,T_{pc}(0) \right),\quad \mathcal{O} \in \{\chi^{\rm BQ}_{11} /\chi^{\rm Q}_{2},~~\chi^{\rm BQ}_{11} /\chi^{\rm QS}_{11}\}
\end{equation} 
shown in Fig.~\ref{fig:ratio_BQ-Tpc}. Lattice continuum estimates reveal enhancements of $\sim2$ for $\chi^{\rm BQ}_{11} /\chi^{\rm Q}_{2}$ (left) and an even more pronounced $\sim2.25$ for $\chi^{\rm BQ}_{11} /\chi^{\rm QS}_{11}$ (right) at $eB\simeq8~{M_{\pi}^2}$, underscoring the potential of $\chi^{\rm BQ}_{11}$ as a magnetometer of QCD. These double ratios are well suited for experiments: they isolate $eB$-induced enhancements while suppressing volume-dependent effects~\cite{STAR:2019ans,ALICE:2025mkk}.

Fig.~\ref{fig:ratio_BQ-Tpc} also shows the corresponding HRG proxy results $R( \sigma_{Q^{\rm PID},p}^{1,1} ~\big/~\sigma_{Q^{\rm PID}}^{2}  )$ (left) and $R( \sigma_{Q^{\rm PID},p}^{1,1} ~\big/~\sigma_{Q^{\rm PID},K}^{1,1}  )$ (right), together with kinematic cut results $R^{cut}_{\rm ALICE/STAR}$ emulating STAR and ALICE detector acceptances. As highlighted in the inset, the proxies retain at least $\sim80\%$ of the magnetic sensitivity predicted by lattice QCD. Even after incorporating kinematic cuts, enhancements of up to 25\% persist at $eB\simeq8~{M_{\pi}^2}$ for $\chi^{\rm BQ}_{11} /\chi^{\rm Q}_{2}$, and up to $60\%$ for $\chi^{\rm BQ}_{11} /\chi^{\rm QS}_{11}$. By bridging theoretical predictions with detector-level analyses, the HRG-based proxies thus offer a viable pathway for probing magnetic-field signatures in HICs through accessible fluctuation observables. These predictions are now being tested experimentally~\cite{ALICE:2025mkk,Nonaka:2023xkg}. The ALICE collaboration has already reported centrality-dependent enhancements in $\chi^{\rm BQ}_{11}/\chi^{\rm Q}_{2}$~\cite{ALICE:2025mkk}, a finding qualitatively consistent with our results. We additionally propose $\chi^{\rm BQ}_{11}/\chi^{\rm QS}_{11}$ as a more sensitive observable, which our results predict to exhibit significantly stronger magnetic enhancement.

\section{QCD EoS in magnetic fields at nonzero baryon density }

The QCD EoS encapsulates the relationships among pressure and energy related bulk thermodynamic observables and their response to the control parameters $(T, eB, \hat{\mu}_{\rm B,Q,S})$~\cite{Ding:2025nyh}. To access nonzero baryon density while circumventing the sign problem, one employs a Taylor expansion in conserved-charge chemical potentials, whose leading-order coefficients are determined by the fluctuations introduced in Sec.~\ref{sec:thm}. In heavy-ion collisions, the three chemical potentials are not independent but are interrelated through strangeness neutrality and isospin asymmetry of the colliding nuclei. At leading order, this yields
\begin{equation}
 \hat{\mu}_{\rm Q}\equiv q_1 (T,eB) \hat{\mu}_{\rm B} + \mathcal{O}(\hat{\mu}_{\rm B}^3),\quad  \hat{\mu}_{\rm S}\equiv s_1 (T,eB) \hat{\mu}_{\rm B} + \mathcal{O}(\hat{\mu}_{\rm B}^3),
\end{equation}
thereby reducing the pressure to a series in $\hat{\mu}_{\rm B}$ alone. The resulting leading-order pressure takes the form
\begin{equation}
    \label{eq:plo}
\hat{P}_{\rm LO} = \frac{1}{2!}\left( \chi^{{\rm B}}_{2} \hat{\mu}_{{\rm B}}^2 + \chi^{{\rm Q}}_{2} \hat{\mu}_{{\rm Q}}^2 +  \chi^{{\rm S}}_{2} \hat{\mu}_{{\rm S}}^2 \right) + \chi^{{\rm B} {\rm Q}}_{11} \hat{\mu}_{{\rm B}} \hat{\mu}_{{\rm Q}} + \chi^{{\rm B} {\rm S}}_{11} \hat{\mu}_{{\rm B}} \hat{\mu}_{{\rm S}} + \chi^{{\rm Q} {\rm S}}_{11} \hat{\mu}_{{\rm Q}} \hat{\mu}_{{\rm S}}\,.
\end{equation}
Upon substituting the constrained chemical potentials, the pressure difference at finite baryon density becomes a pure $\hat{\mu}_{\rm B}$-series~\cite{Bazavov:2017dus,Ding:2025nyh}:
\begin{equation}
    \Delta\hat{P}\equiv \hat{P}(T,eB,\hat{\mu}_{{\rm B}}) -\hat{P}(T,eB,0) =  P_{2}(T,eB)\, \hat{\mu}_{{\rm B}}^{2} + \mathcal{O}(\hat{\mu}_{\rm B}^4)\,,
\label{eq:p2k}
\end{equation}
where $P_{2}(T,eB)$ is the leading-order coefficient implying pressure response to baryon density and explicitly depends on conserved-charge susceptibilities and the chemical potential ratios $q_{1}$ and $s_{1}$.

\subsection{Strangeness neutrality and isospin asymmetry}
\label{subsec:muQovermuB}

In HIC experiments, the colliding nuclei are initially net-strangeness neutral, and their valence quark content constrains the conserved-charge densities: $\hat{n}^{{\rm S}} = 0,\quad  {n}^{{\rm Q}} /  {n}^{{\rm B}} = r$, where $r$ denotes the isospin parameter characterizing the charge-to-baryon ratio. For $^{208}_{82}$Pb$\big/^{197}_{79}$Au nuclei, $r \simeq 0.4$, reflecting slight isospin asymmetry. These constraints, when applied to the leading-order number densities, yield the chemical potential ratios $q_1$ and $s_1$ as explicit functions of the second-order susceptibilities and $r$~\cite{Bazavov:2012vg,Ding:2025nyh}.

\begin{figure}[htbp]
\centering

\includegraphics[width=0.4\textwidth]{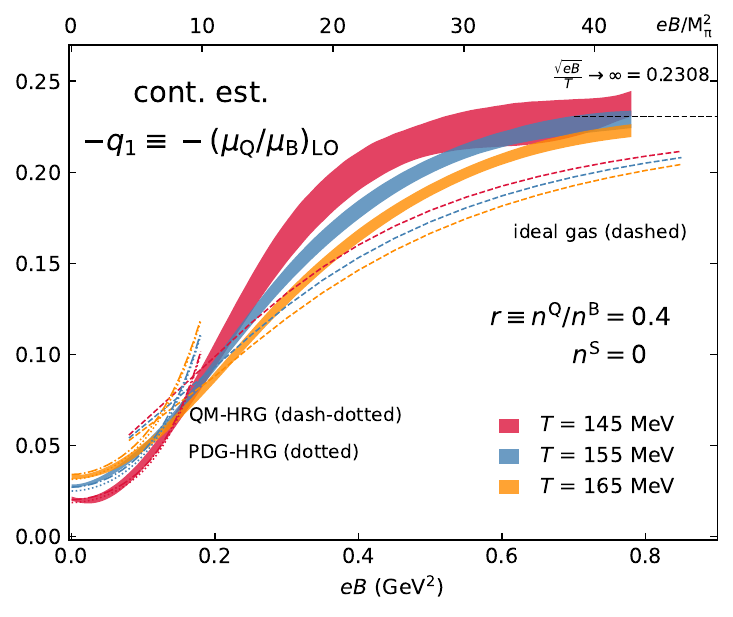}
\includegraphics[width=0.4\textwidth]{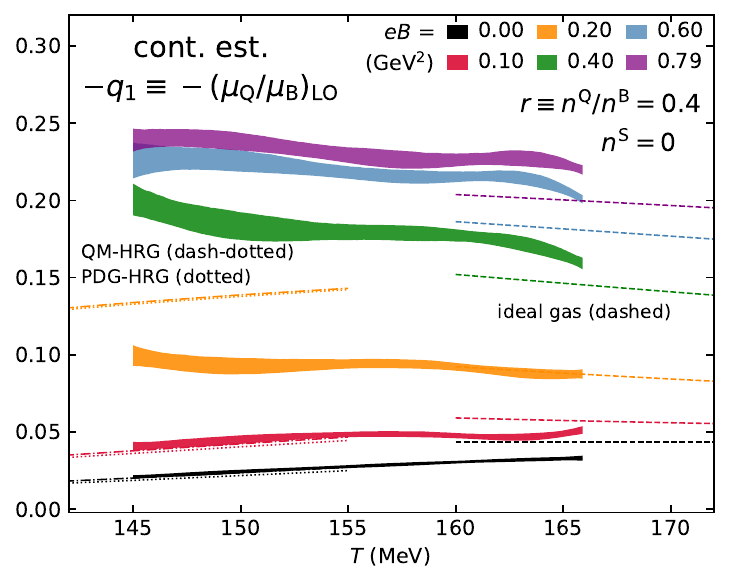}

\includegraphics[width=0.4\textwidth]{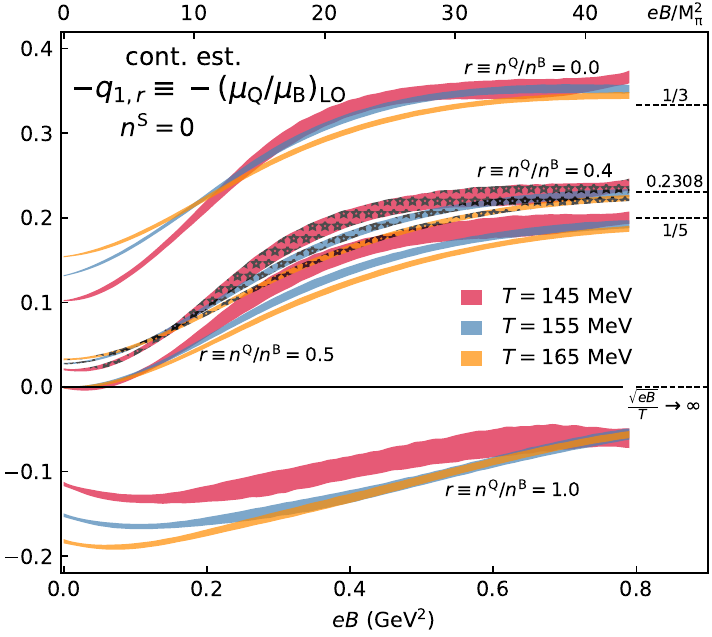}
\includegraphics[width=0.4\textwidth]{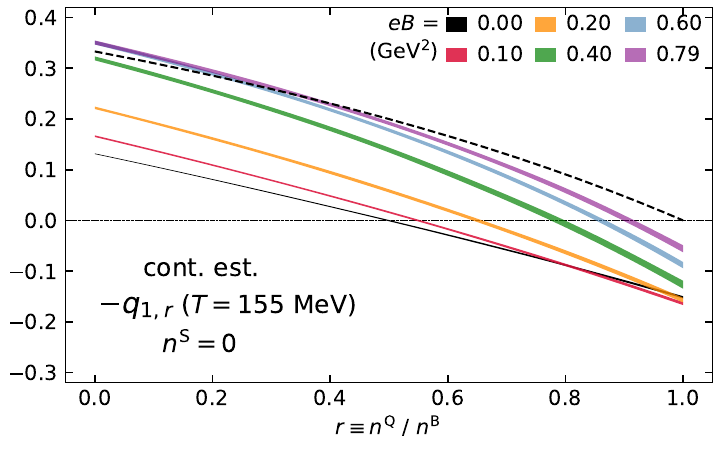}

\caption{Electric charge over baryon chemical potential, $-q_1\equiv -\left(\mu_{\rm Q}/\mu_{\rm B}\right)_{\rm LO}$, for strangeness-neutral systems. {\bf Top}: $eB$- (left) and $T$-dependence (right) for $^{208}_{82}$Pb$\big/^{197}_{79}$Au with $r \simeq0.4$. {\bf Bottom}: $eB$-dependence for various isospin regimes (left) and $r$-dependence for fixed $eB$ (right). Plots are taken from Ref.~\cite{Ding:2025nyh}.}  
\label{fig:q1}
\end{figure}

Fig.~\ref{fig:q1} presents lattice continuum estimates for the leading-order coefficient
\begin{equation}
    q_1(eB,T,r) \equiv \left( \frac{{{{\mu}_{\rm Q}}}}{{{{\mu}_{\rm B}}}} \right)_{\rm LO} = \frac{r\left( \chi_{2}^{{\rm B}}  \chi_{2}^{{\rm S}} - \chi_{11}^{{\rm B} {\rm S}} \chi_{11}^{{\rm B} {\rm S}} \right) - \left(\chi_{11}^{{\rm B} {\rm Q}} \chi_{2}^{{\rm S}} - \chi_{11}^{{\rm B} {\rm S}} \chi_{11}^{{\rm Q} {\rm S}} \right)}{\left(\chi_{2}^{{\rm Q} } \chi_{2}^{{\rm S}} - \chi_{11}^{{\rm Q} {\rm S}}  \chi_{11}^{{\rm Q} {\rm S}}\right) - r\left( \chi_{11}^{{\rm B} {\rm Q}} \chi_{2}^{{\rm S}} -\chi_{11}^{{\rm B} {\rm S}} \chi_{11}^{{\rm Q} {\rm S}} \right)}.
\end{equation}
The top panels correspond to $r=0.4$, relevant for Pb/Au collisions. The top-left and top-right panels display the $eB$- and $T$-dependence, respectively. Lattice results show that $q_1$ is negative across the entire $T$–$eB$ parameter space. Even before introducing magnetic fields, $q_1$ is negative because the isospin asymmetry constraint ($r=0.4$) suppresses the density of positively charged baryons relative to neutral counterparts, requiring a negative $\hat{\mu}_{\rm Q}$. Magnetic fields further intensify this negativity: within the HRG framework, the enhanced degeneracy of the lowest Landau level favors the population of charged baryons, yet the isospin constraint forces $\hat{\mu}_{\rm Q}$ to become even more negative to maintain the required charge-to-baryon ratio. At $eB \sim 0.15~{\rm GeV}^2$, crossings among fixed-temperature continuum bands emerge in the top-left panel, signaling a reversal of the monotonic temperature hierarchy observed at weaker fields. The HRG framework does not capture these crossings, underscoring its limitations in accounting for non-perturbative QCD effects. In the top-right panel, this hierarchy reversal manifests as a sign change in the $T$-slope of $q_1$. At strong fields, lattice results progressively approach saturation, converging toward the magnetized ideal gas limit $-q_1(\sqrt{eB}/T \to \infty) = 0.2308$. This saturation arises from the cancellation of the leading linear $eB$ dependence in the susceptibility ratio defining $q_1$~\cite{Ding:2025nyh, Ding:2025qzh}.

The bottom panels extend the analysis to various isospin parameters $r\equiv n^{\rm Q}/n^{\rm B}\in\left[0,1\right]$, spanning from electrically neutral ($r=0$) to fully charged ($r=1$) baryon systems. The Pb/Au case ($r = 0.4$, hatched bands in the bottom-left panel) is supplemented by the isospin-symmetric case ($r = 0.5$) and the two maximally asymmetric extremes. Since $q_1$ is a ratio observable, all cases saturate to the respective magnetized ideal gas limits at large $eB$. Notably, the isospin parameter $r$ and the magnetic field strength together control the sign and magnitude of $q_1$. This interplay is most clearly reflected in the bottom-right panel, which shows $q_1$ versus $r$ at fixed $eB$. At $eB=0$, a sign change occurs at $r=0.5$; with increasing magnetic field, the sign change shifts to progressively larger $r$ values, and in magnetized ideal gas limit, $q_1$ remains negative throughout the entire $r$ range. 

Similarly, for the strangeness coefficient $s_1\equiv(\mu_{\rm S}/\mu_{\rm B})_{\rm LO}$: magnetic fields enhance the population of charged strange baryons, and strangeness neutrality then requires a slightly more positive $\hat{\mu}_{\rm S}$. The underlying physics parallels the enhancement of $-q_1$, but manifests with the opposite sign because strange quarks intrinsically carry negative strangeness.

\subsection{Leading-order pressure coefficient}
\label{subsec:pressure}

\begin{figure}[t]
\centering

\includegraphics[width=0.4\textwidth]{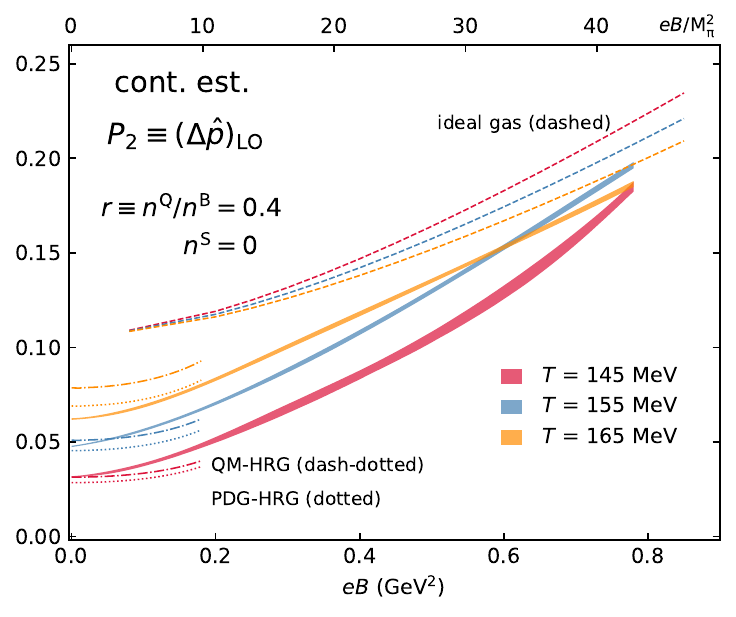}
\includegraphics[width=0.4\textwidth]{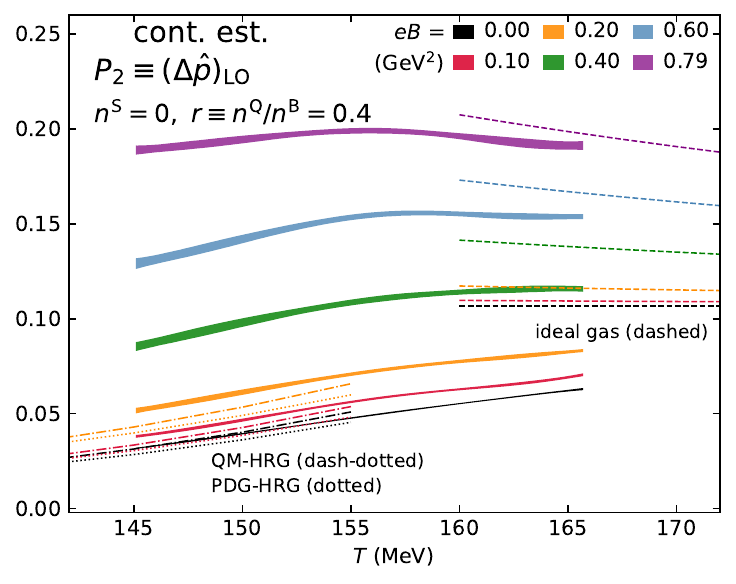}

\includegraphics[width=0.4\textwidth]{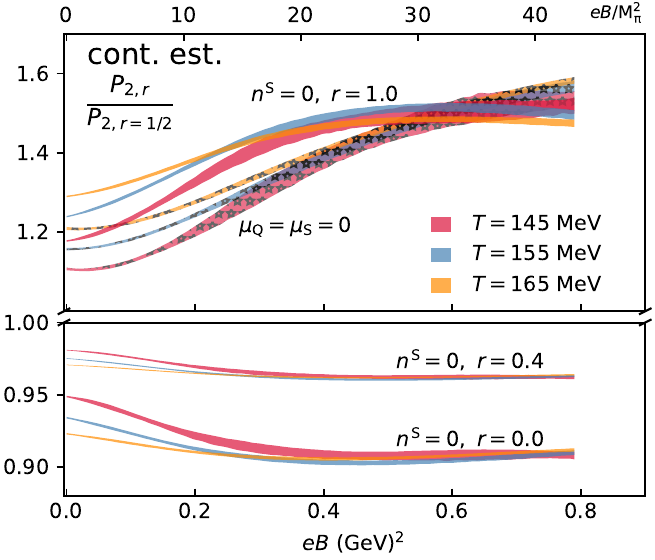}
\includegraphics[width=0.4\textwidth]{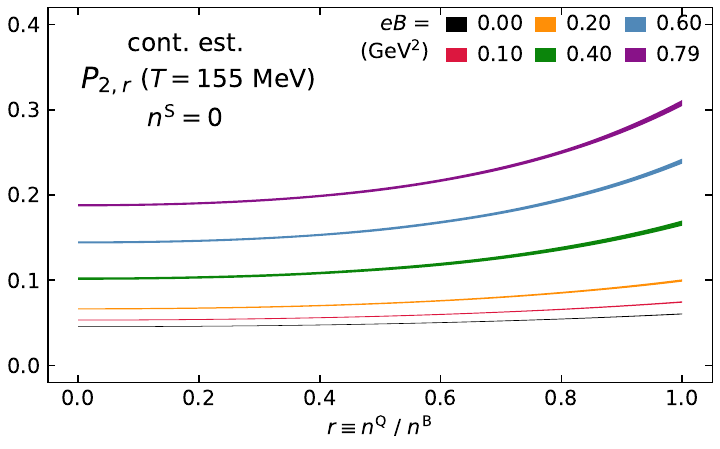}

\caption{Leading-order pressure coefficient $P_2$ for strangeness-neutral systems. {\bf Top}: $eB$- (left) and $T$-dependence (right) for $r=0.4$. {\bf Bottom}: $P_2(r)/P_2(r=0.5)$ versus $eB$ for various isospin regimes (left) and versus $r$ at fixed $eB$ (right). Plots are taken from Ref.~\cite{Ding:2025nyh}.}   
\label{fig:p2_Pb_Au_cont}
\end{figure}

Fig.~\ref{fig:p2_Pb_Au_cont} presents lattice QCD continuum estimates for the leading-order pressure coefficient,
\begin{equation}
P_2(eB,T,r) = \frac{1}{2!}\left( \chi^{{\rm B}}_{2}  + \chi^{{\rm Q}}_{2} q_1^2 +  \chi^{{\rm S}}_{2} s_1^2 \right)  + \chi^{{\rm B} {\rm Q}}_{11} q_1 + \chi^{{\rm B} {\rm S}}_{11}  s_1+ \chi^{{\rm Q} {\rm S}}_{11} q_1 s_1\,,
\end{equation}
which follows from Eq.~(\ref{eq:plo}) upon substituting the constrained chemical potentials. The top-left and top-right panels display the $eB$- and $T$-dependence, respectively. At vanishing magnetic field, $P_2$ increases monotonically with temperature, with a pronounced rise across the QCD transition reflecting the smooth crossover from hadronic to partonic degrees of freedom~\cite{Bazavov:2017dus}. In the presence of magnetic fields, the behaviour becomes significantly more intricate due to the nontrivial interplay between thermal and magnetic effects.

In the top-left panel, $P_2$ increases with $eB$ at all temperatures, driven primarily by the enhanced degeneracy of Landau levels, which scales linearly with the field strength. At relatively weak fields and low temperatures, the HRG model provides a reasonable description of the lattice data. In the strong-$eB$ regime, around $eB\sim 0.6~{\rm GeV}^2$, crossings among fixed-temperature bands emerge, signaling a reordering of the temperature hierarchy and marking a qualitative departure from the monotonic behaviour at weaker fields. This reordering reflects the growing dominance of the lowest Landau level, which governs pressure enhancement through magnetic degeneracy rather than thermal excitation.

In the top-right panel, this reordering manifests as the emergence of non-monotonic structures in the temperature profile. At $eB = 0.79~{\rm GeV}^2$, a mild peak structure becomes evident. The inflection point of $P_2$---typically associated with the pseudo-critical temperature $T_{pc}$---and the emerging peak structures systematically shift toward lower temperatures with increasing $eB$, consistent with the $T_{pc}$-lowering effect observed in previous studies~\cite{Borsanyi:2023buy,Astrakhantsev:2024mat}. At extremely strong fields, lattice results progressively align with magnetized ideal gas predictions (dashed lines), wherein the temperature hierarchy is ultimately reversed. Unlike the ratio observables $q_1$ and $s_1$, the pressure does not saturate with increasing $eB$. This reflects its extensive nature: additional quantum states become accessible through enhanced magnetic degeneracy, driving continued growth.

The bottom panels explore the $r$-dependence of $P_2$ across different strangeness-neutral systems. Like increasing $eB$, increasing the isospin parameter also enhances $P_2$. However, the growth about $r=0.5$ is asymmetric, as shown by the ratio $P_2(r)/P_2(r=0.5)$ in the bottom-left panel. In the strong magnetic field limit, all bands approach saturation. For $r<0.5$, a suppression of at most $\sim10\%$ is observed at $r=0$, while for $r>0.5$, a pronounced enhancement of $\sim 50\%$ appears at $r=1$. This asymmetry reflects the distinct magnetic response of electrically neutral versus charged baryon systems. The bottom-right panel shows the $r$-dependence at fixed $eB$, further illustrating this imbalance. Even stronger enhancements are observed for the unconstrained case $\mu_{\rm Q}=\mu_{\rm S}=0$, corresponding to the pure $\chi^{\rm B}_2$ response.

\section{Summary}
We presented continuum-estimated lattice QCD results for second-order fluctuations of conserved charges and the leading-order QCD equation of state in the presence of strong magnetic fields at nonzero baryon chemical potential, using (2+1)-flavor simulations with the HISQ action at the physical pion mass. Magnetic field strengths were explored up to $eB \simeq 0.8~\text{GeV}^2 \sim 45 M_{\pi}^2$.

The baryon--electric charge correlation $\chi^{\rm BQ}_{11}$ exhibits a striking sensitivity to the magnetic field, with $R_{cp}$-like double ratios reaching enhancements of $\sim2$ ($\chi^{\rm BQ}_{11}/\chi^{\rm Q}_{2}$) and $\sim2.25$ ($\chi^{\rm BQ}_{11}/\chi^{\rm QS}_{11}$) at $eB\simeq8~{M_{\pi}^2}$ along the transition line. Within the HRG framework, proxy observables constructed from final-state hadrons and incorporating kinematic cuts emulating STAR and ALICE detector acceptances retain $\sim80\%$ of the magnetic sensitivity predicted by lattice QCD. The ALICE collaboration has already reported centrality-dependent enhancements in $\chi^{\rm BQ}_{11}/\chi^{\rm Q}_{2}$~\cite{ALICE:2025mkk}, qualitatively consistent with our predictions. We additionally propose $\chi^{\rm BQ}_{11}/\chi^{\rm QS}_{11}$ as a more sensitive experimental observable. 

For the leading-order EoS under strangeness neutrality ($r=0.4$), the chemical potential ratio $q_1\equiv(\mu_{\rm Q}/\mu_{\rm B})_{\rm LO}$ becomes increasingly negative with $eB$ and develops temperature-band crossings at $eB\sim 0.15~{\rm GeV}^2$, signaling a reversal of the monotonic temperature hierarchy not captured by HRG models. The pressure coefficient $P_2$ grows monotonically with $eB$ and, in the strong-field regime ($eB\gtrsim 0.6~{\rm GeV}^2$), develops non-monotonic temperature structures and band crossings consistent with the onset of lowest-Landau-level dominance and the $T_{pc}$-lowering effect. The $r$-dependence of $P_2$ reveals an asymmetric growth about $r=0.5$, reflecting distinct magnetic responses of neutral versus charged baryon systems. 


\paragraph{Acknowledgements}
This work is supported partly by the National Natural Science Foundation of China under Grants No. 12293064, No. 12293060, and No. 12325508, as well as the National Key Research and Development Program of China under Contract No. 2022YFA1604900 and the Fundamental Research Funds for the Central Universities, Central China Normal University under Grants No. 30101250314 and No. 30106250152. The numerical simulations have been performed on the GPU cluster in the Nuclear Science Computing Center at Central China Normal University ($\mathrm{NSC}^{3}$) and Wuhan Supercomputing Center.

\bibliographystyle{JHEP}
\bibliography{refs.bib}

\end{document}